\documentclass[11pt]{article}
\usepackage[margin=1in]{geometry}
\usepackage{graphicx}
\usepackage{url}

\usepackage[caption=false]{subfig}

\usepackage[natbibapa,nodoi,unmask]{apacite}

\title{The Transfer Student Experience: It's A Lot Like Buying a Used Car\thanks{This paper is a prerelease created for the purpose of obtaining feedback. To properly cite this work, please contact the corresponding author in order to obtain the most up-to-date information.}}
\author{Gregory L. Heileman$^{\dagger\natural}$ \and Chaouki T. Abdallah$^{\ddagger}$ \and Andrew K. Koch$^{\flat}$  \\~\\
 heileman@arizona.edu, ctabdallah@gatech.edu, koch@jngi.org\\~\\
 $^\dagger$Department of Electrical \& Computer Engineering \\
  University of Arizona \\ \\ 
  $^\ddagger$Department of Electrical \& Computer Engineering \\
  Georgia Institute of Technology \\ \\
  $^{\flat}$John N. Gardner Institute for Excellence in Undergraduate Education \\ \\
  $^\natural$corresponding author }

\date{}

\begin{document}

\maketitle 

\begin{abstract}
The experience transfer students encounter as they navigate their journeys from community college to university is similar to that of buying a used car. We demonstrate this by showing how the information asymmetry in the market for used cars also occurs in the market for transfer students, producing inefficient markets in both cases, thereby increasing the chances of adverse selection. We diagnose the underlying conditions that produce transfer inefficiencies, identifying them as a structural inequity within the system of higher education. Finally, recommendations for alleviating information asymmetry in transfer processes, that would lead to better outcomes for transfer students, are provided. 
\end{abstract}

{\bf keywords:} transfer articulation, transfer pathways, information asymmetry, structural inequity

\section{Introduction}
More than fifty years ago, economist George Akerlof~(\citeyear{Ak:70}) published a paper that explored how the quality of goods exchanged in a market tends to degrade when information asymmetry exists between buyers and sellers.\footnote{Akerlof~(\citeyear{Ak:01}) recalls that his paper was rejected by two economics journals, one on the grounds that the journal ``did not publish papers on topics of such triviality,'' before it was accepted for publication in the Quarterly Journal of Economics. This work subsequently led to a Nobel Prize~\citep{Nobel:01}. Interestingly enough, Akerlof further recalls that he first recognized the information asymmetry problem in the market for education.} Akerlof considered the market for used cars as an example, and he demonstrated that because buyers are generally not able to distinguish between ``cream puffs'' and ``lemons,'' they are only willing to pay a price that corresponds to the average value between these two extremes. This tends to drive away the sellers of high-quality cars who are unwilling to sell their cars below their true value, leaving more low-quality cars behind in the marketplace, which in turn produces a disincentive for manufacturers to produce high-quality cars. Thus, the information asymmetry around car quality between buyers and sellers results in an inefficient marketplace for used cars.  

Research published in the decade following Akerlof's study suggests that information asymmetry in the used car marketplace may be particularly problematic for low-income consumers and consumers of color. Specifically, \cite{McNeTrMi:79} showed that low-income buyers paid ``more for used cars, got less redress for defects discovered after purchase, and were less satisfied and more likely to believe something was misrepresented.'' The trend reported by McNeil and colleagues is alive and well more than forty years later, and the dynamic is further troubling when examining the race and ethnicity of car buyers. Low income and minority consumers are far more reliant upon the used car market place~\citep{Pa:03}. \cite{Va:19}, in a report published by the National Consumer Law Center, noted that many race and ethnicity disparities in the car buying market arise ``because the market for cars is troublingly opaque and inconsistent.'' He concludes, a ``more consistent and transparent marketplace would not only benefit consumers of color but all marketplace participants.''

The information asymmetry and inequitable consumer dynamics that has existed in the market for used cars are strikingly similar to problems encountered by transfer students in higher education.  Specifically, we can think of transfer articulation as a marketplace, where students are the buyers, colleges and universities are the sellers, and transactions involve the seller applying the buyer's prior academic work towards the satisfaction of degree requirements. In the current transfer articulation marketplace, sellers have more and better information than buyers about the expected quality of these transactions. In the case of used cars, information asymmetry leads to an inefficient marketplace that puts buyers at a distinct disadvantage in the relationship and increases the chances of \emph{adverse selection}. In the case of the transfer articulation marketplace, the adverse selection issue is clear, students suffer financial loss if they choose a lemon program that recognizes little prior work, particularly if they are unable to realize this until \emph{after} they have ``bought'' the new college or university experience. Moreover, the inefficiency of the transfer market encourages predatory practices by low-quality for-profit institutions~\citep{Bo:19,BoRoCoShMe:18,Sh:14}.

Similar to the used car market, low-income and minority students are far more reliant upon the transfer articulation market place. In particular, low-income and minority students are more likely than the college-going population as a whole to begin their higher education journeys at a community college~\maskcitep{RoKo:20}. Thus, information asymmetry in transfer articulation translates into market inefficiencies that lead to significant equity issues. Indeed, transfer processes produce some of the most inequitable outcomes in all of higher education. In 2018 there were approximately 17 million undergraduate students attending degree granting postsecondary institutions in the United States, with roughly 6 millions of these students enrolled in community colleges~\citep{HuEtAl:20}. Of these students, based upon historical data, we can expect about 35\% of the total population will transfer at least once and 11\% twice during their academic careers. In doing so, they will on average lose the equivalent of one year of course work with each transfer~\citep{Si:14}. With the average annual cost of college tuition at \$3,500 for community colleges, and \$10,000 for universities, a lost year equates to an annual excess tuition of more than \$50 billion. When lost opportunity costs related to wages, retirement savings, and student debt are factored in, the loss associated with transfer inefficiencies can be conservatively estimated to exceed \$150 billion per annum~\citep{Do-Ga:16}. To put this loss into perspective, consider that it is more than three times the average annual cost of damages due to weather and climate disasters in the United States over the past forty years~(NOAA, \citeyear{NOAA}). It should also be recognized that the financial costs are only one part of the overall loss. The societal impact of the dashed hopes, crushed dreams, and unmet aspirations of those who simply walk away from higher education because of transfer inefficiencies is easy to imagine, but difficult to quantify. Because these adverse outcomes are disproportionately borne by low-income and minority students, we identify transfer articulation as a significant structural inequity built into the system of higher education in the United States. 

The fact that transfer processes are massively inefficient is well recognized. Indeed, colleges, universities, statewide systems, and even state legislators have all recognized the problem, and many have proposed, or mandated, solutions aimed at alleviating the situation~\citep[see][]{FrAn:21}. Unfortunately, however, most of these efforts have had minimal impact, as the statistics provided above clearly indicate.  In this article, we consider the subtle and often misunderstood reasons that give rise to the significant inefficiencies in transfer articulation, and we discuss some of the measures that can be taken to alleviate adverse transfer outcomes. 

\section{Why is Transferring so Hard?}
With good intentions, many universities work with their community college partners to create transfer articulation pathways that provide a roadmap for students moving from one program, e.g., A.S. Biology, at a community college to a similar program, e.g., B.S. Biology, at a four-year institution. A quick perusal of university websites will lead you to these transfer articulation plans, where you will often find that they are either out-of-date, or the particular program you might be interested in transferring to has not been considered. Furthermore, in reviewing these plans, we often find errors.  The reasons for these problems are not due to sloppiness or lack of concern. Rather, what is often not understood is that the combinatorics associated with transfer articulation quickly overwhelm any attempts to solve the problem via the brute-force methods that are commonly applied; that is, by bringing together faculty members from the departments at the community college and university to create transfer articulation plans. When these labor-intensive and time-bound efforts inevitably falter, the response often involves throwing more advising human resources at the problem, but the material impact of these efforts is often negligible. And students, particularly students of color and students experiencing poverty, bear the brunt of the negative consequences implicated in the transfer system's inefficient and opaque design. 

To see why a brute-force approach is doomed to failure, consider a typical state system involving six public universities and fifteen community colleges. Assume that each of these institutions has 100 programs, and we wish to create transfer pathways from each of the community colleges to each of the universities. Let us even simplify the problem by limiting the number of transfer pathways from each program at a community college to only five programs at each university. Then a lower bound on the total number of pathways from the in-state community colleges to each university is: 
\[
 15 \ \mbox{CCs} \times 100 \ \mbox{CC programs} \times 5 \ \mbox{university programs} = 7,500 \ \mbox{pathways},
\]
where CC denotes a community college.  If we consider the number of pathways statewide, there are:
\[
 7,500 \ \mbox{pathways} \times 6 \ \mbox{universities} = 45,000 \ \mbox{total pathways statewide}.
\]

\noindent
If only one hour is spent on each of these pathways, $3.75$ years of a person's full-time effort would be required to create these $7,500$ pathways for a single university, and the total effort to create pathways for the entire state would involve more than 22 person-years of effort. Furthermore, this only accounts for in-state transfer students, and many fail to recognize that it is only the first layer of complexity associated with this problem. 

A more significant issue is the fact that transfer articulation plans are generally created under the assumptions that a student enters the associate degree program without any prior credits, that the associate degree will be completed first, and that no curricular changes in the bachelor's degree program will occur while the student is pursuing the associate degree. In reality, students often have a diversity of prior credits from multiple sources, they change their degree aspirations, and curricula are in constant flux as they are updated and improved as a part of continuous quality improvement efforts. Attempting to create transfer pathways that account for these additional variabilities layers on significant additional complexities. When we also account for the actual coursework associated with each individual student, and the time it takes to determine course equivalences, it is easy to see that this problem quickly becomes unmanageable, putting students at a distinct disadvantage regarding the information needed to make transfer-related decisions. These pathways also hide further complexities that often only come to light once a student attempts to transfer, as is described in more detail next.

\subsection{Creating a Completion Plan}
In order to decide among the myriad transfer opportunities just described, ideally a student would be able to fully understand the work that would be required to complete a particular degree program, with a full accounting of how the credits they previously earned would apply at another institution. However, this is in fact the root of the information asymmetry problem alluded to previously. A transfer student typically only has a vague notion of the amount of work that will be required to complete a degree through transfer, and a clear pathway to completion is often not revealed until \emph{after} they have committed to a transfer institution; that is, after they have ``purchased'' the transfer. It is not uncommon for this to involve a revelation that very little prior coursework transferred to the receiving institution in a meaningful way; in other words, they come to the realization that they bought a lemon. This is problematic for all students; especially those who have limited financial resources and a fixed set of terms for which they will receive need-based financial aid. 

\begin{figure}[h]
 \centerline{\includegraphics[width = 4in]{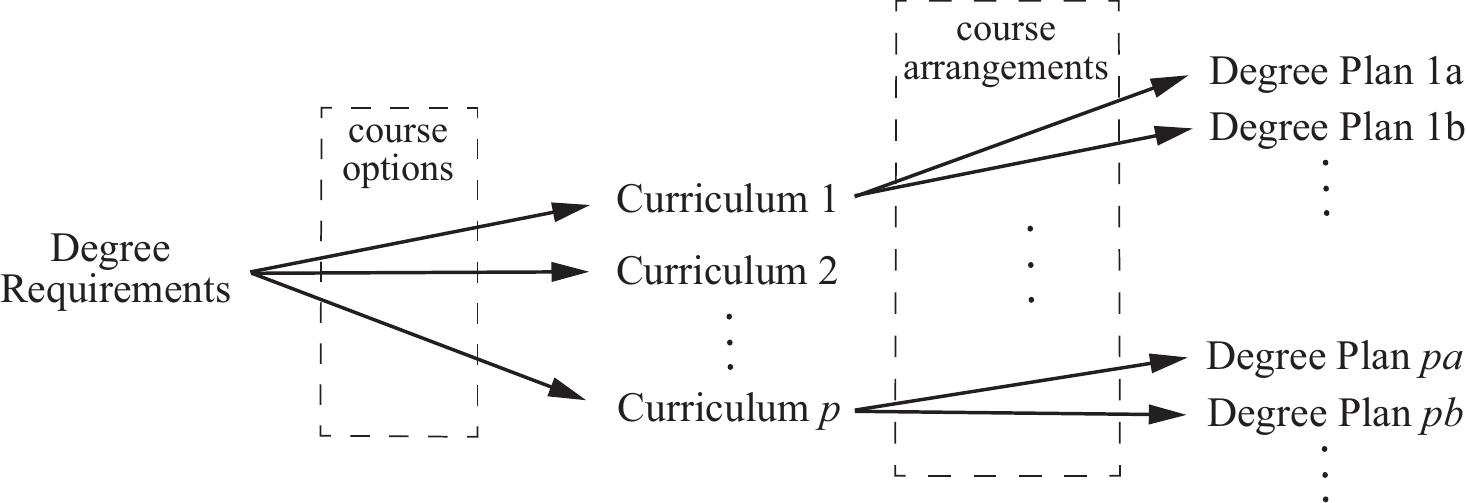}}
 \caption{The relationships between degree requirements, curricula, and degree plans.}\label{data-types}
\end{figure}

The inefficient and unjust market for transfer articulation we have just described is generally not due to unscrupulous universities acting like ``used car salesmen.'' Rather, the inefficiency has more to do with the complexity of the problem, and the fact that university information systems do not readily provide the most essential pieces of information in a useable form. To fully appreciate this, we must ``look under the hood'' in order to better understand the mechanics of transfer articulation, and to do this, one must first appreciate the distinctions between \emph{degree requirements}, \emph{curricula}, and \emph{degree plans}, as depicted in Figure~\ref{data-types}. 
These three terms are often confounded when considering issues related to transfer articulation, further contributing to the information asymmetry problem. Specifically, college and university advisors generally have a much better understanding of the nuances that accompany these terms, and how they impact transfer and completion, than do transfer students and their families, as described next.

\subsection{Important Terminology}
In order to earn the credential associated with an academic program, e.g., B.A. Fine Art, a student must satisfy all of the degree requirements associated with that academic program. In this sense, degree requirements, shown on the far left in Figure~\ref{data-types}, are the most vital information for a student to appreciate, as they determine whether a student earns a degree or not. A significant problem is that degree requirements can be quite complicated to express, and even more complicated to understand. Specifically, degree requirements are almost exclusively satisfied by earning sufficient grades in the specific courses that constitute each degree requirement. For example, undergraduate programs typically contain a ``General Education'' degree requirement. This requirement often contains sub-requirements such as a ``Mathematics,'' satisfied by passing a certain number of courses from a specific list of allowable math courses, etc. An entire set of degree requirements for an academic program can be represented as a Boolean formula, consisting of many sub-formula that correspond to these degree requirements and sub-requirements. For instance, a simple version of a General Education degree requirement might be represented as:

{\small 
\[
 Gen \; Ed \; Core \; =  \;  (Math \; Core)  \; \wedge  \; (Humanities \; Core)  \; \wedge  \; (Writing \;Core)  \; \wedge  \; (Fine \; Art \; Core)  
\]}

\noindent
where $\wedge$ denotes logical conjunction, and each of the four items shown inside parentheses correspond to the sub-requirements of the General Education degree requirement. In other words, all four sub-requirements must be satisfied in order to satisfy the General Education requirement as a whole. 

The mathematics component in the General Education requirement, i.e., $Math \; Core$, might be satisfied by either passing a particular math course, or by demonstrating prior learning on a placement exam, e.g., 

{\small 
\[
 Math \; Core \; =  \; (Statistics) \; \vee  \; (Precalculcus)  \; \vee  \; (Calculus \; I)  \; \vee  \; (placement \; exam)  
\]}

\noindent
where $\vee$ denotes logical disjunction. The first three terms on the right hand side of the equality above can be satisfied by earning a sufficient grade in the corresponding class, and the $placement \; exam$ term can be satisfied by earning a sufficient grade on any one of a variety of placement exams~(i.e., it is also a disjunctive sub-formula). Thus, the Math Core is satisfied by satisfying any one of its sub-requirements.

A realistic set of degree requirements may contain on the order of one hundred such sub-requirements, and some of these sub-requirements might be satisfied by taking some combination selected from a list of one hundred or more different courses. This leads to a combinatorial explosion in terms of the number of possible ways to satisfy a set of degree requirements through course-taking options, and this is on top of the excessive number of options a student has for choosing a transfer pathway described previously. For most programs, we can conservatively estimate that hundreds of thousands of different curricula can be constructed to satisfy a single set of degree requirements for a degree program. In Figure~\ref{data-types} we denote this by showing that a single set of degree requirements can lead to many different curricula depending upon the course options selected. This complexity serves to highlight the importance of academic advisors, who routinely work with students to select a collection of courses, i.e., a curriculum, that will allow them to satisfy degree requirements as efficiently as possible.  We have also developed a framework for quantifying how the complexity of these curricula impact a students ability to progress through them \maskcitep{HeAbSlHi:18}.

Even if a student decides on the set of courses in a curriculum, combinatorics makes a final troubling appearance when they try to construct a degree plan from a particular curriculum.  We have seen that a curriculum is just a collection of courses that, if successfully completed, satisfies a set of degree requirements. A degree plan, then, is a term-by-term arrangement of the courses in a curriculum constructed to satisfy the prerequisite relationships between the courses in the curriculum. The fact that a single curriculum can lead to many different degree plans, depending upon how the courses are arranged, is depicted on the right side of Figure~\ref{data-types}. For instance, if we consider a bachelor's degree curriculum consisting of forty courses, there are over one million different ways to arrange these courses over eight terms when there are no prerequisites. By accounting for prerequisites, we might reduce the possibilities by a factor of ten or more, but we are still talking about thousands of possible degree plans at the very least. A logical question one might then ask is, ``Are some of these degree plans better than others?'' Here again, the role of academic advisors is critical. Certainly, for some students, one degree plan is far better than others in assuring student success. For example, experienced advisors routinely counsel against taking particular courses together in the same term that may be toxic in combination for certain students. For instance, some students might have better outcomes by taking Calculus~I followed by Physics~I, rather than by taking them together in the same term, as is common practice in some science and engineering programs. Creating a degree plan that spans multiple institutions, and involves earning multiple credentials, only complicates this work.

\subsection{The Mechanics of Transfer}
\begin{figure}
 \centerline{\includegraphics[width = 6in]{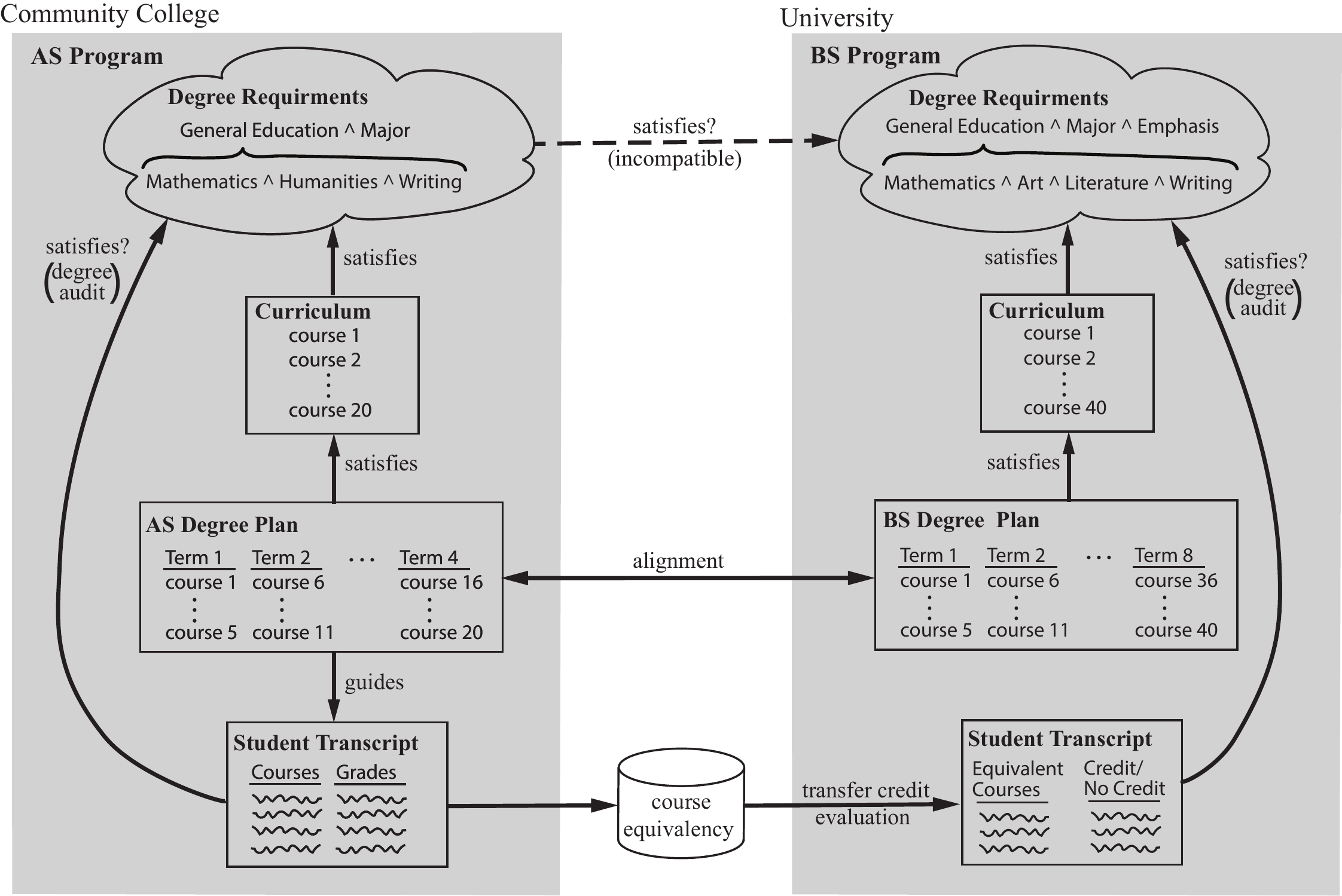}}
 \caption{The high-level mechanics of transfer involving a community college and a university, detailing the structures that exist within the institutions and how they connect.}\label{transfer-mechanics}
\end{figure}

\begin{figure}
 \centerline{\includegraphics[width = 6.75in]{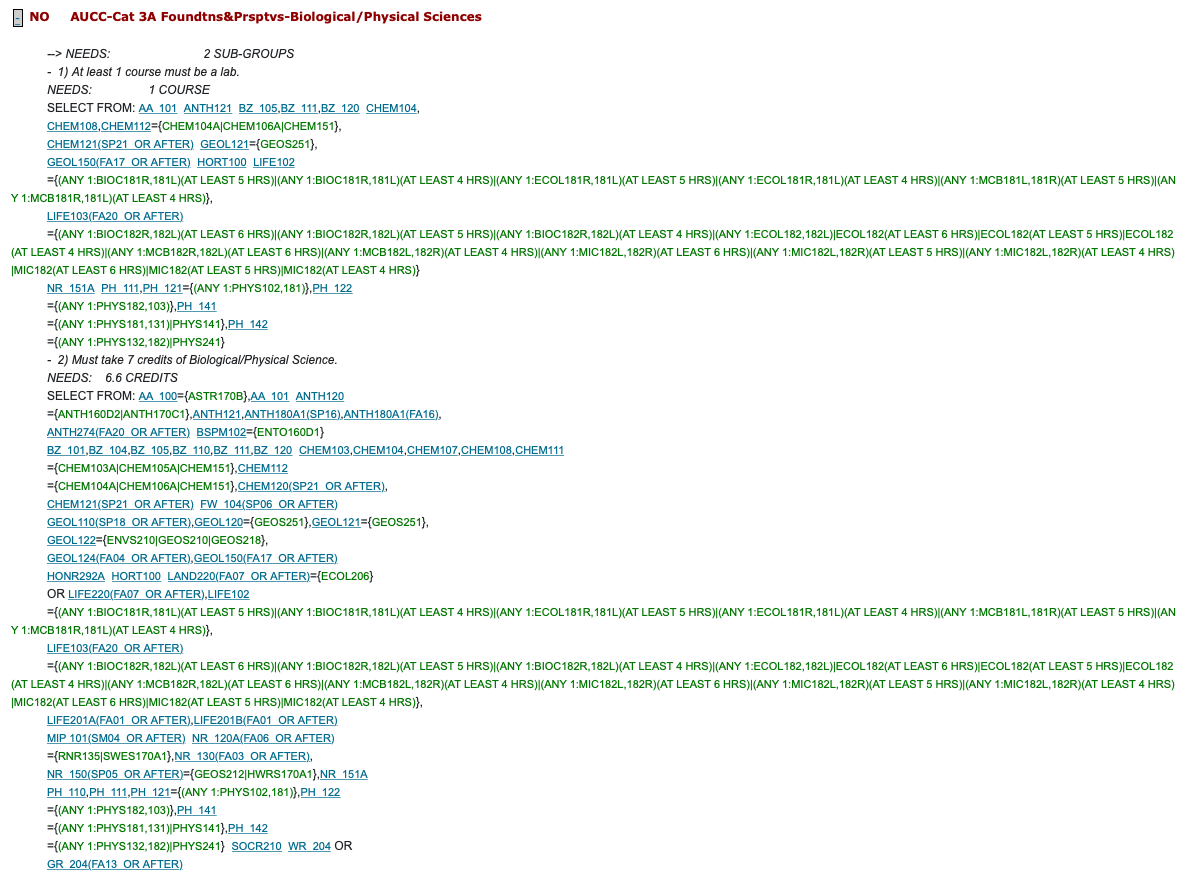}}
 \caption{One of the 35 ``rules'' that must be satisfied as a part of an actual degree audit at a university. This rule, which has not yet been satisfied, as indicted by the ``NO'', involves two sub-rules that can be satisfied in a multitude of ways.}\label{audit-example}
\end{figure}

With these definitions in mind, we are now in a position to describe the detailed mechanics of transfer articulation, as well as the common places where this machinery fails. In Figure~\ref{transfer-mechanics} we map the processes involved in transfer across the two institutions involved when a community college student transfers to a university. 
In this figure, the processes within the community college are shown inside the shaded panel on the left, and those within the university are shown inside the shaded panel on the right. Within each institution, the relationships between degree requirements, curricula, and degree plans are as described above; however, what is typically presented to the students at each institution are web pages that contain helpful degree plans for the degrees being pursued. That is, students are generally only vaguely aware that many different curricula exist for a degree program, and they are likely even less aware of the underlying degree requirements associated with the degree program. The key point is that within each institution, degree plans have been carefully constructed so that if students follow them, they will earn their degrees. Thus, at the bottom of the Community College panel in Figure~\ref{transfer-mechanics} we note that hopefully the courses that show up on a student's transcript were taken using the guidance provided by the degree plan. In order to verify that this coursework in fact satisfies the degree requirements, an institution typically performs a \emph{degree audit} in order to certify that a student is eligible to graduate. An excerpt extracted from an actual eleven-page audit is shown in Figure~\ref{audit-example}. 
This figure contains the criteria for satisfying a single portion of a general education requirement, namely Biological/Physical Sciences. Note that this sub-requirement itself contains two sub-requirements, one of which contains two disjunctive Boolean clauses, and these clauses themselves contain additional disjunctive and conjunctive Boolean clauses. To say that this type of degree audit presentation is in any way ``student friendly'' is an obvious understatement, yet we are aware of numerous institutions providing similar degree audits to students during advising sessions.

\subsubsection{Information System Incompatibilities}
Now consider the situation of a student attempting to transfer from the AS program to the BS program shown in Figure~\ref{transfer-mechanics}. Note the dashed arrow drawn between the two sets of degree requirements depicted in this figure. Ideally, it would be possible to query these in order to determine whether particular degree requirements in the community college program might also satisfy degree requirements in the university program; however, as shown in the figure, this is generally not possible due to various incompatibilities. For instance, the systems used to store degree requirements at the two institutions may be from different vendors, the courses used to satisfy various degree requirements may not exist at both institutions, and the information systems themselves are probably not set up to allow for information sharing. Thus, although an advisor at the community college may work with a student to ensure they are taking courses that satisfy the AS degree requirements, it is exceedingly difficult to also check that they are also satisfying the underlying BS degree requirements at the university. 

The only relevant university information typically accessible to transfer students and their advisors are the degree plans provided on university websites. In Figure~\ref{transfer-mechanics} we show how the transfer student might work to align their AS degree plan with some BS degree plan published on a university website. However, this effort often lacks guarantees that the courses will transfer, and more importantly, that they will satisfy BS degree requirements. 

\subsubsection{Credit Recognition Versus Application}
In order to obtain a formal certification of how their credits will apply, a transfer student generally needs to submit their transcript to a university as a part of an application process. As shown at the bottom of Figure~\ref{transfer-mechanics}, this involves creating a transcript for the transfer student at the receiving institution through a transfer credit evaluation. Specifically, if a course offered by the community college has substantially similar learning outcomes to some course offered by the university, a student successfully completing the community college course can petition to have this course accepted as transfer credit by the university. The work involved in establishing these so-called course equivalences typically involves a review of the community college course syllabus by a faculty member in the department that offers the potentially equivalent course at the university. This is often a time-consuming process that can take months to complete and is sometimes referred to as credit recognition. Unfortunate bias can also be introduced at this stage. For instance, we have heard university faculty claim that the community college version of their course is not ``good enough'' to allow it to be transferred into their esteemed program. When you consider that universities receive thousands of these requests per year, it is not surprising to observe significant processing backlogs in the offices that manage these requests.  Finally, once this work is completed, it becomes possible to perform a degree audit over the transfer coursework in order to determine how much of it actually applies.

Many who are unfamiliar with the intricacies of transfer articulation, believe the entire transfer problem is solved through large scale credit recognition; that is, by universities accepting as equivalent many of the courses offered by local community colleges. However, as we are now able to fully explain, nothing could be further from the truth. For instance, state legislatures often create bills related to transfer, such as requiring common course numbering across all institutions of higher education in the state~(e.g., \cite{AZ:12,NM:18}). Indeed, we have heard some professionals within the transfer articulation community claim this legislation often simply adds more administrative burden, doing more harm than good in terms of creating clarity for transfer students. There also exists entities at both the state and national levels that work to establish transfer equivalences between institutions~(e.g., \cite{AZTransfer,Transferology}). Although these efforts are important, when using these tools, transfer students are often surprised to learn that their prior credits transfer, but they end up counting as excess credit hours at the receiving institution. That is, although the credits are recognized, they cannot be used to satisfy any degree requirements in a program at the receiving institution. Thus, even though a state legislator may believe the problem is solved once legislation is created requiring everyone to accept credits from one another within the state, the credits may transfer in such a way that they are rendered useless at the receiving institution. For instance, transfer courses often end up counting as ``elective credits'' on a student's transcript. In this case, there is little difference between the credit transferring and not transferring, in either case they are not be applied towards requirements satisfaction at the receiving institution, and therefore they do not move the student any closer to earning a bachelor's degree. We refer to this as the \emph{credit application problem}, and it is what makes all the difference to transfer students in terms of their ability to efficiently complete a bachelor's degree.

\subsubsection{An Example}
The credit application problem is so pervasive yet misunderstood in higher education that we refer to it as the ``dirty little secret'' of transfer articulation. Those ``in the know,'' such as transfer advisors at colleges and universities, have a common mantra to describe this problem, ``it's not that the credits count, it's how they apply.'' Transfer advisors are not maliciously hiding this dirty little secret from students, they just do not have the tools they need to bring the necessary facts to light. To illustrate the problem, let us consider the case of Julia, who is active duty in the Air Force and was previously stationed at Kirtland Air Force Base in Albuquerque, New Mexico. While located there, she earned credits related to her interests in computing from Central New Mexico Community College, namely the first two programming courses in a typical computer science curriculum, as well as a course on the mathematical foundations of computing. When she was subsequently reassigned to Davis-Monthan Air Force Base in Tucson, Arizona, she took an additional course at Pima Community College, namely a course on computer networking. Now that she has completed her tour of duty, she is interested in moving back to her home in Colorado, where she is interested in using her GI benefits to earn a bachelor's degree in computer science at some university in her home state. 

\begin{figure}
 \centerline{\includegraphics[width = 5.5in]{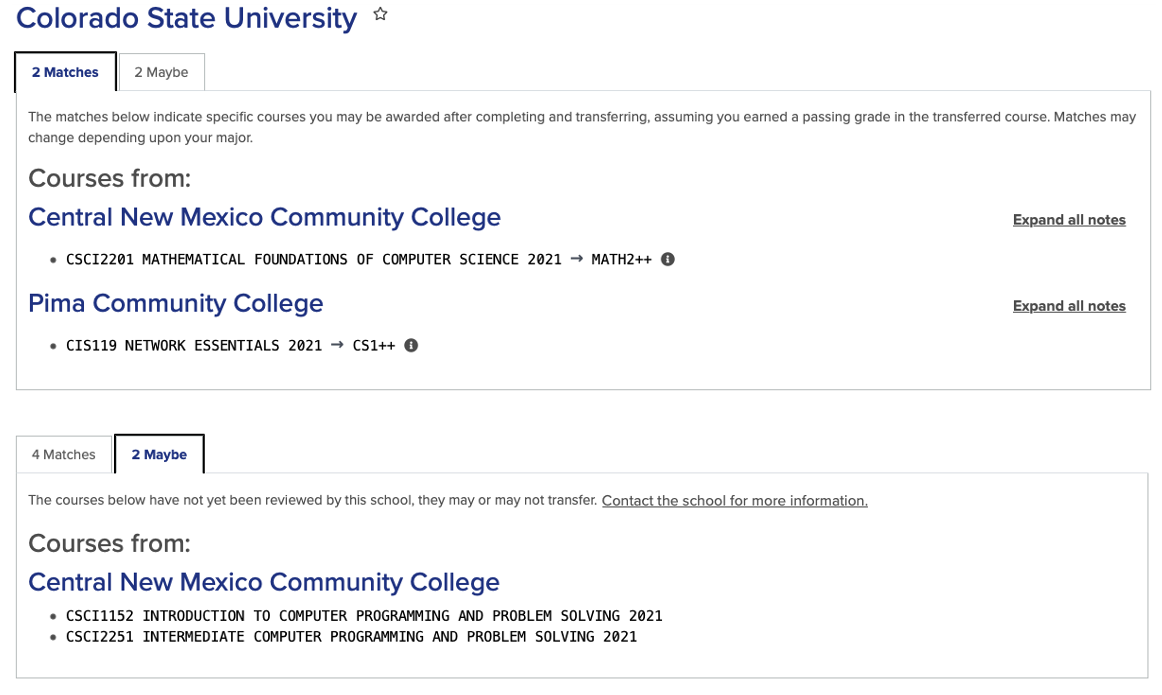}}
 \caption{An example transfer credit analysis for a hypothetical student take from the Transferology website. Of the six prior courses, only four will currently transfer, and the credits accepted for this prior coursework will only transfer as elective credits.}\label{xfer-example}
\end{figure}

To narrow down the choices in her transfer decision, Julia consulted the Transferology.com website, where she input her prior coursework.\footnote{Transferology is a commercial transfer student portal containing nationwide data, see: transferology.com.}  The analysis that was returned for one potential transfer destination, Colorado State University, is shown in Figure~\ref{xfer-example}. 
This analysis shows that Colorado State University, which yielded the highest ``match'' on this website, will immediately recognize two courses for transfer credit, and the other two courses have not yet been evaluated for any type of transfer equivalency. It is also important to note how the two courses that do transfer will be recognized by Colorado State University; both would transfer as lower-level \emph{electives}. Thus, although these courses would show up as transfer credits on a Colorado State University transcript, there is a good chance they cannot be used to satisfy any degree requirements. 

Given that two of the courses have not been evaluated by Colorado State University, Julia's next steps might involve formally requesting their evaluation. There is a reasonable likelihood these courses will articulate in a way that satisfies some of the degree requirements in a computer science program. Indeed, if Julia were to consult with an academic advisor at Colorado State University, there is some likelihood that even the courses that will transfer as electives would be reconsidered; that is, they might be allowed to satisfy some degree requirements within the computer science program. 
  
Thus, the situation for Julia at Colorado State University is perhaps better than the initial analysis seems to indicate; however, the amount of effort she must exert to make her prior coursework count is significant, and it requires persistent self-advocacy. If you take into account that a similar level of effort would be required at every other university she is interested in transferring to, it is easy to see that the amount of work confronting a typical transfer student considering multiple transfer destinations is overwhelming.  

\section{Mitigating Information Asymmetry in Car and Transfer Markets}
Efficient markets are built upon a framework of information-symmetry among market participants. If this condition is violated, with some participants having more information than others, inequitable transactions are more likely to occur, leading to a loss of trust in the market itself. When these inequities are concentrated on particular disadvantaged populations, they become an inequity built into societal structure; that is, they constitute a \emph{structural inequity}. In the case of the market for used cars, much has been done since the 1970's to mitigate structural inequity, and we can certainly learn from them in addressing inequities in the transfer student market.  

\subsection{Expert Verification and History}
A major contributing factor to information asymmetry in the used car market is the fact that cars are highly complex mechanical structures, making it exceedingly difficult for non-experts to assess their quality. One method buyers have for dealing with this is through \emph{expert verification}, namely, by having a trained mechanic inspect a car prior to purchasing it.

We have seen that the transfer process is also exceedingly complex; however, the complexity in this case is primarily due to the combinatorics involved. It is far more difficult, at the moment, to conduct independent expert verification of the quality of transfer articulation pathways, as opposed to assessing the quality of a used car. The good news, however, is that unlike car inspections, a transfer pathway inspection came be made into an efficiently computable process. However, this will require the adoption of a standard representation for degree requirements, along with large-scale public disclosure of program degree requirements using this representation. An issue we discuss in more detail below.

The used car marketplace has also recently been disrupted by numerous services that have worked to diminish information asymmetry around the quality of used cars by revealing their history. Specifically, large data sets now exist containing detailed information about individual vehicles such as sales history, recall and warranty information, insurance claims, and accident history~\citep[e.g.,][]{Autocheck}. Numerous online applications now pair this data with other helpful information such as price predictions and projected depreciation, in order to put the buyer in a much better position to assess the quality of used cars~\citep[e.g.,][]{Carfax,Cargurus,Carvana}. This has led to a much more efficient marketplace for used cars. We contend that it is also time for a similar disruption in the market for transfer students in higher education. Specifically, much could be learned by making outcomes data available for students who previously pursued various transfer pathways. For instance, it would be useful for transfer students to have access to information detailing how credits have been applied in the past within the particular programs they are considering, how many students were able to complete the transfer pathway, and how much time and money was required to complete the pathway.

\subsection{Guarantees and Warranties}
Another way to alleviate inequities in markets is to provide warranties or guarantees about the quality of the goods being purchased. For instance, car dealers now routinely offer extended warranties for used cars, and many aftermarket vendors also sell warranties. However, in the used car market, the effectiveness of these products in building trust is undercut by the numerous scams perpetrated by unscrupulous actors~\citep{FTC}.  

Somewhat related guarantees provided to transfer students in many states (e.g.,~\cite{UC-Xfer,UCF-Xfer}). Specifically, many universities have programs that involve providing guarantees of admission to student attending particular community colleges, as long as the maintain a certain level of academic performance. Here again, the community college student may have very limited visibility into how their prior credits will apply at the receiving institution. Thus, to fully realize the potential of such programs, universities must work to not only guarantee admission, but to also to apply prior earned credits towards the satisfaction of bachelor's degree requirements. 

\subsection{Governments and Law}
A final important analogy can be drawn to the used car market in the area of legislation. Over time, governments have worked to protect consumers by enacting legislation that holds car manufacturers liable for allowing ``lemons'' to enter the market place. These so-called ``Lemon Laws'' serve to protect buyers even after a sale has been completed~\citep{Ha:21}. Thus, they provide assurances that work to build trust, decrease information asymmetry, and therefore improve the efficiency of the used car market. 

In the case of the transfer student market, legislation aimed at not just ensuring that credits transfer, but that they also apply, is needed. Given that faculty ``own'' the curricula at their respective institutions, such legislation must be carefully crafted to ensure faculty participation, so that the quality of academic programs can be maintained. For instance, legislation aimed at improving the \emph{visibility} of how transfer credits apply would significantly reduce information asymmetry, and would go a long way towards helping transfer students make informed choices. This approach also supports the creation of markets-based solutions. Universities that do a poor job of applying transfer credits would be clearly revealed, thereby providing an incentive for them to improve the situation if they hope to effectively compete for transfer students.

\section{Making Transfer More Transparent and Equitable}
In this article we have made the case that the information asymmetry problem present in transfer articulation is largely due to the combinatorial complexities involved, the time-consuming nature of credit recognition, and the opaqueness of credit applicability. As we have described it, the arcane and convoluted nature of transfer articulation procedures effectively buries important information deep within bureaucratic university ``machinery'' that is difficult to access and understand in a timely manner. Thus, transfer students are often still working to figure out how to make their prior credits count long after they have already transferred. This is particularly deleterious to students who lack both the financial capital to fund extra terms of study, as well as the cultural capital needed to navigate complex educational systems, making this a structural inequity in higher education.

Although the transfer articulation problem is complex, it is important to recognize that it is an efficiently computable problem. Given a student record, courses equivalences, and the degree requirements of an academic program, we can compute the precise degree requirements satisfied by a student's prior credits, as well as a completion plan for earning a degree using this prior coursework. Indeed, if degree requirements are available in a reliable form, numerous additional capabilities become possible, all serving to create a more efficient transfer student marketplace. Two important use cases immediately come to mind. First, using the computational capabilities we have just described, it becomes possible for college and university administrators to create two-year-to-four-year transfer plans. As we have already described, these transfer roadmaps are notoriously difficult to create and maintain.  Typically, the starting point for these plans are the degree plans for the associates and bachelor's degrees in a single discipline, rather than the underlying degree requirements. Thus, the resulting transfer roadmaps end up being less robust than they could be. Furthermore, by computing transfer analyses on-the-fly from available information, it becomes possible to construct any pathway from a two-year program to a four-year program. For instance, a student could easily map out a pathway from an associate degree in business to a bachelor's in psychology, which is something not generally supported when creating standardized roadmaps ``by hand.'' 

Another important use case enabled by the computational capabilities described above involves the ability for transfer students to perform what-if analyses over the various transfer scenarios available to them. We contend that this capability would essentially eliminate the information asymmetry that currently exists around the transfer articulation problem. And eliminating information asymmetry in transfer articulation can be directly correlated with increased equity in transfer, as it has been shown to be in the marketplace for cars~\citep{MoZeSi:01}. Specifically, by giving transfer students the ability to see exactly how their prior coursework will apply towards the satisfaction of particular degree requirements in specific programs, the guesswork is taken out of the transfer decision. Indeed, we envision online applications similar to those now available in the used car marketplace described above. For example, using the aforementioned computational capabilities, one could easily construct an application that evaluates a student's transcript relative to all of the bachelor's programs in a given state system, and then provides an analysis showing which of these programs the student is closest to completing, along with the costs to complete each program. With this knowledge, fully informed decisions regarding transfer can be made, leading to more satisfied buyers in the marketplace for transfer articulation, along with more equitable outcomes for transfer students.

Using these capabilities, we can also envision the creation of a transfer navigation app that fully supports information symmetry. With current real-time route navigation apps, e.g., Google Maps, a new route to the desired destination can be computed in real-time whenever a wrong turn is made, or a new destination is selected, accompanied by the non-famous ``recalculating route'' message.  A similar capability should be provided to transfer students well before their date of actual transfer. Specifically, the ability to query how a particular set of courses at a community college satisfies the degree requirements at a given university will enable the construction of real-time degree pathway navigation tools. If a community college student completes a course, or fails to do so, or changes their mind on the major they would like to pursue, they should be able to quickly and easily visualize a ``recalculated route'' to the bachelor's degree that accounts for transfer credit articulation. We reiterate that the capabilities we have just described only become possible when two conditions are met. First, an open data standard for representing degree requirements must be created and widely accepted. Second colleges and universities must publicly release their current degree requirements in this format so that others may query them.

\begin{figure}
 \centerline{\includegraphics[width=5in]{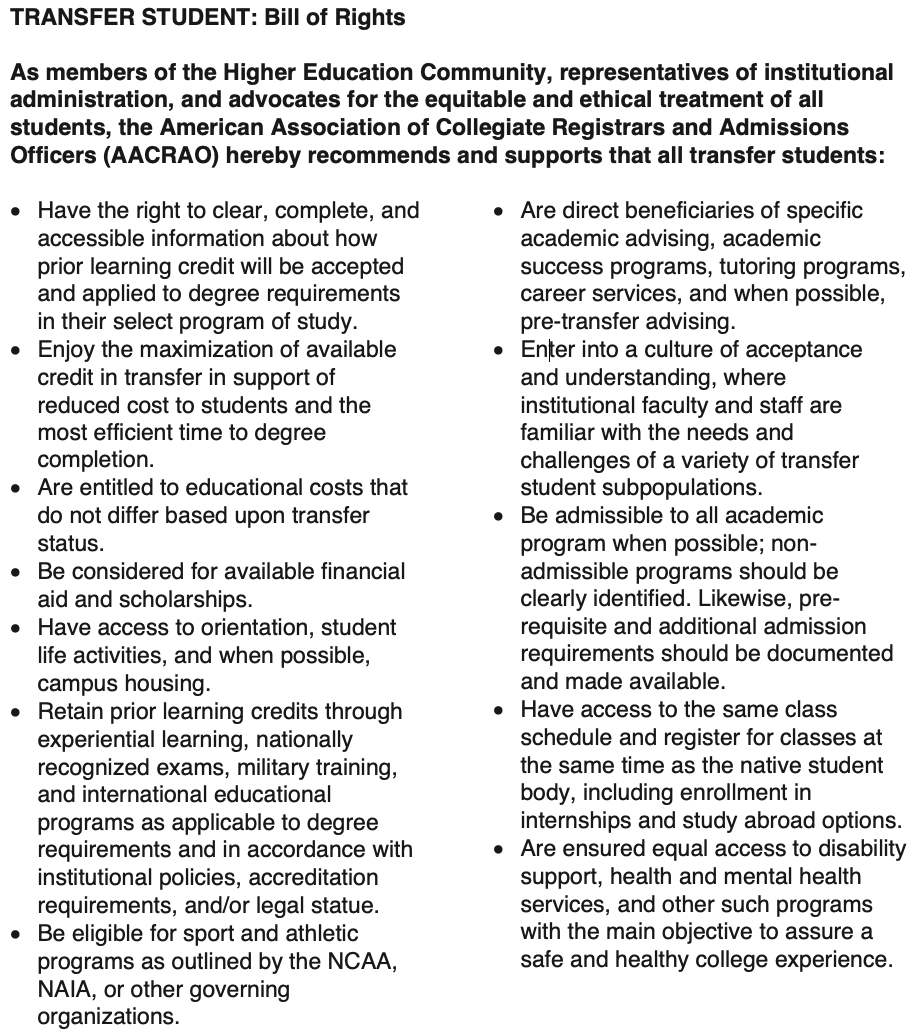}}
 \caption{AACRAO Transfer Student Bill of Rights (AACRAO, \citeyear{AACRAO:17}).}\label{AACRAO}
\end{figure}

The American Association of Collegiate Registrars and Admissions Officers (AACRAO) organization recently published a Transfer Student Bill of Rights that calls for great transparency in transfer processes (see Figure~\ref{AACRAO}). 
The first two articles in this Bill of Rights squarely address the importance of supporting timely decision-making, as we have also described here. As is the case with self-help programs, recognizing the problem is the all-important first step; doing something about it is what comes next. And the good news is, registrars nationwide are in a position to fully appreciate the complexities associated with this problem, and to perhaps do something about it. Indeed, they may be better positioned than any group in higher education in this regard. In Figure~\ref{transfer-mechanics} we noted the key missing element needed to support more efficient decision-making as a part of transfer articulation, namely, the ability to reason over the satisfaction of degree requirements across institutional boundaries. This can only happen with the establishment of universally accepted standards for representing degree requirements, along with a willingness to make this information publicly available so that others may query it. Registrars are typically responsible for curating this information at their institutions, and are thus well positioned to lead the way in making it publicly available.


\newpage

\bibliography{transfer}
\bibliographystyle{apacite}

\end{document}